\documentclass[letterpaper, 10 pt,  conference]{ieeeconf}  % Comment this line out
\usepackage{graphicx} % for pdf, bitmapped graphics files
\usepackage{amsmath,bm,times} % assumes amsmath package installed
\usepackage{amssymb}  % assumes amsmath package installed
\usepackage{tikz}
\usetikzlibrary{shapes,arrows,backgrounds,fit,positioning}
\usepackage{subfigure}
\usepackage{cite}
\usepackage{soul}
\usepackage{balance}
\usepackage{tabularx}
\usepackage{url}
\allowdisplaybreaks

\newtheorem{assumption}{Assumption}
\newtheorem{problem}{Problem}
\newtheorem{proposition}{Proposition}
\newtheorem{remark}{Remark}
\newtheorem{theorem}{Theorem}

\newtheorem{lemma}{Lemma}
\newtheorem{example}{Example}

\newcommand{\qed}{$\hfill\blacksquare$}

\DeclareMathOperator*{\argmax}{arg\,max}

\title{\bf Payoff Mechanism Design for Coordination \\ in Multi-Agent Task Allocation Games}% 
\author{Shinkyu Park and Julian Barreiro-Gomez% <-this % stops a space
\vspace{-0.5cm}
  \thanks{Park's work was supported by funding from King Abdullah University of Science and Technology (KAUST). Barreiro-Gomez's work was supported by the Center on Stability, Instability, and Turbulence (SITE) and 
  Tamkeen under the NYU Abu Dhabi Research Institute grant CG002.}
  \thanks{Park is with the Electrical and Computer Engineering, King Abdullah University of Science and Technology (KAUST), Thuwal 23955, Saudi Arabia. {\tt shinkyu.park@kaust.edu.sa}} \\
  \thanks{Barreiro-Gomez is with NYUAD Research Institute, New York University Abu Dhabi, PO Box 129188, Abu Dhabi, United Arab Emirates. {\tt jbarreiro@nyu.edu}}
}

\IEEEoverridecommandlockouts
\begin{document}

\maketitle
% \thispagestyle{empty}
% \pagestyle{empty}

%%%%%%%%%%%%%%%%%%%%%%%%%%%%%%%%%%%%%%%%%%%%%%%%%%%%%%%%% 
\begin{abstract}
  We investigate a multi-agent decision-making problem where a large population of agents is responsible for carrying out a set of assigned tasks. The amount of jobs in each task varies over time governed by a dynamical system model. Each agent needs to select one of the available strategies to take on one or more tasks. Since each strategy allows an agent to perform multiple tasks at a time, possibly at distinct rates, the strategy selection of the agents needs to be coordinated.
  We formulate the problem using the population game formalism and refer to it as the \textit{task allocation game}. We discuss the design of a decision-making model that incentivizes the agents to coordinate in the strategy selection process.
  
  As key contributions, we propose a method to find a payoff-driven decision-making model, and discuss how the model allows the strategy selection of the agents to be responsive to the amount of remaining jobs in each task while asymptotically attaining the optimal strategies. Leveraging analytical tools from feedback control theory, we derive technical conditions that the model needs to satisfy, which are used to construct a numerical approach to compute the model. We validate our solution through simulations to highlight how the proposed approach coordinates the agents in task allocation games.

\end{abstract}

%%%%%%%%%%%%%%%%%%%%%%%%%%%%%%%%%%%%%%%%%%%%%%%%%%%%%%%%% 
\section{Introduction} \label{sec:introduction}

We investigate task allocation games to study coordination in repeated strategic interactions in a large population of agents.
Consider that there is a finite number of tasks to be carried out by the agents. We quantify the amount of jobs remaining in each task with a positive variable, and every agent can select one of the available strategies at a time to take on one or more tasks. The main objective
is to design a decentralized decision-making model that allows the agents to coordinate and
minimize remaining jobs in all tasks.

Task allocation games are relevant to engineering applications. For instance, in multi-robot resource retrieval \cite{9561809, 681242}, a team of multiple robots is tasked with searching and collecting target resources across partitioned areas in a given environment. Each task can be defined as collecting resources from an area and the strategy selection refers to taking one of the tasks. In target tracking applications \cite{6558009}, a group of mobile units with heterogeneous sensing capabilities are deployed to collect data about the states of multiple targets of our interests. Based on the type of equipped sensors, each mobile unit can collect different sets of data on the targets' states. A task is defined as collecting data on a portion of the target states and a strategy specifies which pair of available sensors a mobile unit can equip. In both scenarios, the amount of resources to collect and the data to gather vary depending on past strategy selection of the agents and also on environmental changes and target dynamics.

To design a model for the agent strategy selection in such engineering applications, we investigate task allocation in dynamically changing environments. Multi-agent task allocation problems have been widely studied across various research communities \cite{1677945, 8439076, 5674046, 5072249, 10.1007/978-3-030-20915-5_54, 5161293, doi:10.1080/00207179.2016.1231422}. A game-theoretic approach to the problem using replicator dynamics is investigated in \cite{10.1007/978-3-030-20915-5_54}. The authors of \cite{8439076, 5674046} use the hedonic game to study the coordination of multiple agents in task allocation. Applications of population game approaches to address task allocation in swarm robotics \cite{5161293} and the control of a water distribution system \cite{doi:10.1080/00207179.2016.1231422} are discussed. Also, relevant to the task allocation game that we investigate in this work, whose formalism is defined in a state space, the state-based potential game has been studied in \cite{MARDEN20123075}, and the design of state-based game to solve distributed optimization is proposed in \cite{6459524}.

A majority of existing works assume that the environment underlying the game is static and aim to find the optimal task allocation. In contrast, 
we study the design of a decision-making model under which the agents can repeatedly switch among multiple tasks to minimize remaining jobs in the tasks.
We adopt the population game formalism \cite{Sandholm2010-SANPGA-2} to state the problem and to study the decision-making model design. The model prescribes how the agents take on a given set of tasks and how the agents should switch among the tasks by revising their strategy selection to asymptotically attain optimality. 
We consider that each agent in the population is given a set of $n$ strategies to carry out assigned tasks where we denote the agents' strategy profile -- the distribution of the agents' strategy selection -- by a non-negative vector $x = (x_1, \cdots, x_n)$. Remaining jobs associated with $m$ tasks are denoted by a non-negative vector $q = (q_1, \cdots, q_m)$ for which a dynamic model describes how $q$ changes -- both growth by environmental changes and reduction by the agents -- based on the agents' strategy selection $x$.  

Based on the evolutionary dynamics framework \cite{Sandholm2010-SANPGA-2}, we specify a decentralized decision-making model that allows individual agents to revise their strategy selection based on a payoff vector $p = (p_1, \cdots, p_n)$, where each $p_i$ is the payoff an agent receives when it selects the $i$-th strategy. As the main contribution,
we design a payoff mechanism to define how $p$ should depend on $q$ to encourage the agents to select the tasks with more jobs to perform and asymptotically attain the minimum of a given cost $c(q)$.
Applying convergence analysis tools \cite{EDM_water_passivity,Fox2013Population-Game,Franklin_jbg,9029756,park2018cdc, 9219202} that are based on passivity theory in the population games literature, we establish conditions under which the agents' strategy profile converges and asymptotically attain the optimal profile. We use the conditions to compute the payoff mechanism.

The paper is organized as follows. In Section~\ref{sec:problem_description}, we explain the task allocation game formulation and the main problem we address in this paper. In Section~\ref{sec:incentive_design}, we present the main result on the payoff mechanism design and analysis on convergence of the agent strategy revision process to the optimal strategy profile. In Section~\ref{sec:simulations}, we present simulation results 
to illustrate our main contribution. We conclude the paper with a summary and future plans in Section~\ref{sec:conclusions}.

\section{Problem Description} \label{sec:problem_description}
Consider a large population of agents that are assigned with $m$ tasks and are given $n$ strategies to carry out the tasks.\footnote{The number of tasks is not necessarily the same as that of available strategies, i.e., $m \neq n$.} We associate each task $j \in \{1, \cdots, m\}$ with a variable $q_j \geq 0$ which quantifies the amount of jobs remaining in the task. Let $x_i \geq 0$ denote the portion of the agents selecting strategy~$i \in \{1, \cdots, n\}$ and a fixed positive number $M$ to be the mass of the agent population satisfying $M = \sum_{i=1}^n x_i$.\footnote{Considering the population state $x$ as control input to 
\eqref{eq:q_dynamics}, the population mass $M$ can be interpreted as a limit on the control input.} Each agent selects one of the strategies at a time based on payoff vector $p = (p_1, \cdots, p_n)$.

Let $\mathbb R_+^n$ be the set of all $n$-dimensional vectors with non-negative entries, and let $\mathbb X_M$ be the space of all feasible states $x = (x_1, \cdots, x_n)$ of the population defined as
  $\mathbb X_M = \left\{ x \in \mathbb R_+^n \,\big|\, \textstyle \sum_{i=1}^n x_i = M \right\}.$
Given a matrix $G \in \mathbb R^{n \times m}$, we represent $G$ using its column and row vectors as follows:
\begin{align} \label{eq:G_representation}
  G = \begin{pmatrix} G_1^{\tiny \text{col}} & \cdots & G_m^{\tiny \text{col}} \end{pmatrix} = \begin{pmatrix} G_1^{\tiny \text{row}} \\ \vdots \\ G_n^{\tiny \text{row}} \end{pmatrix}.
\end{align}

\subsection{Task Allocation Games}
To investigate the task allocation problem, we formalize the problem as a large population game in which the agents select strategies to perform jobs in the assigned tasks quantified by $q = (q_1, \cdots, q_m)$. The vector $q$ varies over time based on the agents' strategy selection and changes in the environment. Hence, each agent needs to evaluate and adaptively select a strategy based on $q$.

Given $x(t)$ and $q(t)$, at each time $t$, the following ordinary differential equation describes the rate of change of $q(t)$.
\begin{align} \label{eq:q_dynamics}
  \dot q(t) &= - \underbrace{\mathcal F(q(t), x(t))}_{\text{reduction rate}} + \underbrace{w}_{\text{growth rate}}, ~ q(0) = q_0 \in \mathbb{R}^m_+,
\end{align}
where $\mathcal F : \mathbb R_+^m \times \mathbb R_+^n \to \mathbb R_+^m$ is a continuously differentiable mapping\footnote{To have the reduction rate mapping defined for any population mass $M$, we define the domain of $\mathcal F$ as $\mathbb R_+^m \times \mathbb R_+^n$.} that defines the \textit{reduction rate}, which quantifies how fast the agents adopting strategy profile $x$ reduce $q$, and the constant vector $w = (w_1, \cdots, w_m) \in \mathbb{R}^m_+$ represents the \textit{growth rate} for $q$ due to environmental changes. To ensure that the positive orthant $\mathbb R_+^m$ is forward-invariant for $q(t)$ under \eqref{eq:q_dynamics}, 
each $\mathcal F_i$ of $\mathcal F = (\mathcal F_1, \cdots, \mathcal F_m)$ satisfies $\mathcal F_i(q, x) \leq w_i \text{ if } q_i = 0$. For notational convenience, let us define $\mathbb O$ as the set of stationary points of \eqref{eq:q_dynamics}, i.e., 
\begin{align} \label{eq:q_dynamics_stationry_points}
    \mathbb O = \{(q,x) \in \mathbb R_+^m \times \mathbb X_M \,|\, \mathcal F(q,x) = w \}.
\end{align}

We make the following assumption on the mapping $\mathcal F$.

\begin{assumption} \label{assumption:on_F}
  The reduction rate $\mathcal F_i$ for each task~$i$ depends only on its associated variable $q_i(t)$ and the agent strategy selection $x(t)$, and increases as does $q_i(t)$. For instance, in the resource retrieval application discussed in Section~\ref{sec:introduction}, when there is a larger volume of resources spread out across the areas, the robots would need to travel a shorter distance on average to locate and retrieve the resources and hence given a fixed strategy profile $x$, the variable $q_i(t)$ decreases at a faster rate. We formalize such assumptions as  $\frac{\partial \mathcal F_j}{\partial q_i} (q,x) = 0$ if $i \neq j$ and $\frac{\partial \mathcal F_i}{\partial q_i} (q,x) > 0$. \hfill $\square$
\end{assumption}

According to Assumption~\ref{assumption:on_F}, we represent the reduction rate as $\mathcal F(q, x) = (\mathcal F_1 (q_1, x), \cdots, \mathcal F_m (q_m, x))$,
where for fixed $x$, each $\mathcal F_i$ is an increasing function of $q_i$.

\begin{remark} \label{eq:unique_q}
  Suppose that given $x$ in $\mathbb X_M$, there is $q$ in $\mathbb R_+^m$ satisfying $\mathcal F(q,x) = w$. By Assumption~\ref{assumption:on_F}, $q$ is unique. \hfill $\square$
\end{remark}

The following examples illustrate how the dynamic game model \eqref{eq:q_dynamics} can be adopted in control systems applications.
\begin{example} [Multi-Robot Resource Collection \cite{9561809}] \label{example:resource_collection}
  Let $m = n$ and $\mathcal F = (\mathcal F_1, \cdots, \mathcal F_n)$ be defined as
  \begin{align} \label{eq:q_dynamics_example_1}
    \mathcal F_i(q_i, x_i) = R_i \frac{\exp(\alpha_i q_i) - 1}{\exp(\alpha_i q_i) + 1}x_i^{\beta_i},
  \end{align}
  where $R_i$, $\alpha_i$, and $\beta_i$ are positive constants. 
  The parameter $R_i$ represents the maximum reduction rate associated with strategy~$i$, and $\alpha_i$ and $\beta_i$ are coefficients specifying how the reduction rate $\mathcal F_i$ depends on $q_i$ and $x_i$, respectively. Note that each function $\mathcal F_i$ satisfies $\mathcal F_i(0, x) = 0$ and Assumption~\ref{assumption:on_F}. Here, $m = n$ and 
  only the agent selecting strategy~$i$ can reduce
  $q_i$ associated with task~$i$.
\end{example}

\begin{example} [Heterogeneous Sensor Scheduling \cite{6558009}] \label{example:heterogeneous_sensing}
    We adopt the model \eqref{eq:q_dynamics} as an abstract description of how mobile units' sensor scheduling affects the uncertainty reduction in estimating states of multiple targets. Let $m < n$ and $\mathcal F = (\mathcal F_1, \cdots, \mathcal F_m)$ be defined as
  \begin{align} \label{eq:q_dynamics_example_2}
    \mathcal F_i(q_i, x) = \sum_{j \in \mathbb N_i} R_i \frac{\exp(\alpha_i q_i) - 1}{\exp(\alpha_i q_i) + 1} x_j^{\beta_i},
  \end{align}
  where $\mathbb N_i$ denotes the set of the strategies (available sensor configurations of a mobile unit) that can collect data on the state of the $i$-th target. The parameters $R_i$, $\alpha_i$, $\beta_i$ have the same interpretation as in Example~\ref{example:resource_collection}. Unlike the previous example,
  the strategies are defined to allow the agents to reduce multiple task-associated variables of $q$.
\end{example}

\begin{example}[Water Distribution Control \cite{EDM_water_passivity,7823106,Franklin_jbg}]
There are $m$ reservoirs each of which is assigned with a respective maximum water level $\bar l_i$ for $i$ in $\{1, \cdots, m\}$.
Denote by $(l_1 (t), \cdots, l_m(t))$ water levels of the reservoirs at each time $t$.
Let $w$ be a constant outflow specifying water demands by consumers and $(x_1(t), \cdots, x_n(t))$ be controllable inflows, where $n$ does not necessarily coincide with the number $m$ of reservoirs. The simplified dynamics for the water levels can be defined as $\dot l_i(t) = \mathcal{F}_i(\bar l_i - l_i(t),x(t)) – w$ satisfying 
$\mathcal{F}_i(0,x)=0$ to ensure that each reservoir cannot hold water above its maximum level:
for instance, $\mathcal{F}_i(\bar l_i - l_i, x) = \frac{(\bar l_i - l_i)}{\bar l_i} x_i$. By defining 
$q_i(t) =\bar{l}_i – l_i(t)$ as
remaining space in each reservoir~$i$,
we can derive the dynamic model as 
\begin{align*}
   \dot q_i(t) = - \frac{1}{\bar l_i} q_i(t) x_i(t) + w.
\end{align*}
\end{example}

\subsection{Agent Strategy Revision Model} \label{sec:strategy_selection_model}
Our model is based on the evolutionary dynamics framework \cite{Sandholm2010-SANPGA-2} in which the \textit{strategy revision protocol}  $\varrho_{i}^\theta: \mathbb R^n \to \mathbb R_+$ determines an agent's strategy revision based on the payoff vector $p \in \mathbb R^n$, where $\theta = (\theta_1 , \cdots, \theta_n) \in \mathbb X_M$ is a parameter of the protocol. We adopt the Kullback-Leibler Divergence Regularized Learning (KLD-RL) protocol \cite{9483414, park2023learning} to define 
$\varrho_{i}^\theta(p)$ as
\begin{align} \label{eq:kld_rl_protocol}
  \varrho_{i}^\theta(p) = \frac{\theta_i \exp(\eta^{-1} p_i)}{\sum_{l=1}^n \theta_l \exp(\eta^{-1} p_l)},
\end{align}
where $\eta > 0$. The protocol $\varrho_{i}^\theta(p)$ describes the probability of an agent switching to strategy~$i$ given $p$ and $\theta$.
Note that the smaller the value of $\eta$, the more the strategy revision depends on the value of $p$.

Each agent is given an opportunity to revise its strategy selection at each jump time of an independent and identically distributed Poisson process, and uses the protocol to select a new strategy or keep its current strategy selection. Since the strategy revision of individual agents only depends on the payoff vector and takes place independently of each other, their decision-making is decentralized and the coordination among them occurs implicitly through their decision-making model. Based on discussions in \cite[Chapter~4]{Sandholm2010-SANPGA-2}, as the number of agents in the population tends to infinity, the following ordinary differential equation describes how each component of $x(t) = (x_1(t), \cdots, x_n(t))$ evolves over time.
\begin{align} 
  \dot x_i(t) &= \mathcal V_i^\theta (p(t), x(t)) \nonumber \\
              &= \textstyle \sum_{j=1}^n x_j (t) \varrho_{i}^\theta (p(t)) - x_i(t) \textstyle \sum_{j=1}^n \varrho_{j}^\theta (p(t)).\label{eq:edm}
\end{align}
We refer to \eqref{eq:edm} as the Evolutionary Dynamics Model (EDM).

Note that at an equilibrium state $(p^\ast, x^\ast)$ of the EDM \eqref{eq:edm} under the KLD-RL protocol \eqref{eq:kld_rl_protocol}, if $\theta = x^\ast$, the following implication holds:
\begin{align} \label{eq:equilibrium_state}
  x_i^\ast > 0 \implies p_i^\ast = \max_{1 \leq j \leq n} p_j^\ast.
\end{align}
Eq. \eqref{eq:equilibrium_state} means that every agent receives the highest payoff at $(p^\ast, x^\ast)$ if the parameter $\theta$ of \eqref{eq:kld_rl_protocol} is the same as $x^\ast$.

\begin{figure}
  \center
  \subfigure[]{
    \includegraphics[trim={.1in .1in .1in 0in}, clip, width=1.40in]{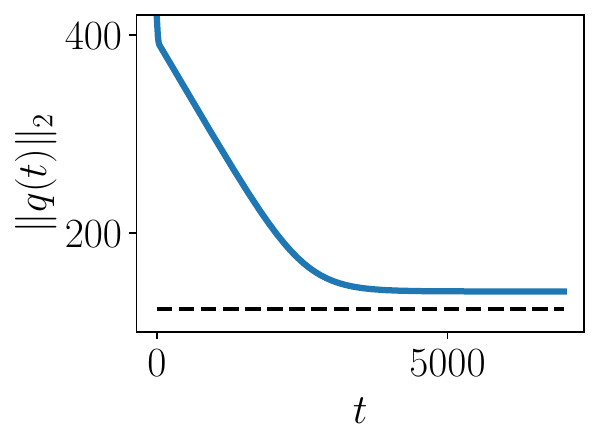}
    \label{fig:cost_example}
  }
  \subfigure[]{
    \includegraphics[trim={.1in .1in .1in 0in}, clip, width=1.47in]{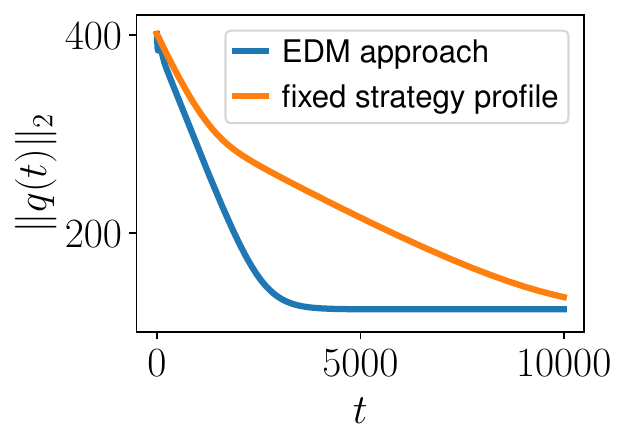}
    \label{fig:comparison_strategy_revision}
  }
  
  \caption{Graphs depicting trajectories $\|q(t)\|_2, \, t \geq 0$ determined by \eqref{eq:q_dynamics} and \eqref{eq:edm} using Example~\ref{example:resource_collection}. The parameters of \eqref{eq:q_dynamics} are defined as $m=n=4$, $M=1$, $R_i = 3.5$, $\alpha_i = 0.05$, $\beta_i = 1$, $w = (0.05, 0.25, 1.00, 2.00)$, and those of \eqref{eq:kld_rl_protocol} as $\theta = x^\ast$, $\eta = 0.001$, where $(q^\ast, x^\ast) \in \mathbb O$ is the equilibrium state minimizing $\max_{1 \leq i \leq 4} q_i$. In (a), the dotted black line is the minimum $2$-norm achievable when the payoff mechanism $p=Gq$ is optimally designed, and the blue line represents the trajectory $\|q(t)\|_2, \, t \geq 0$ when the population state is determined by \eqref{eq:edm} with $p=q$. In (b), the blue and orange lines represent the trajectories $\|q(t)\|_2, \, t \geq 0$ when the population state is determined by \eqref{eq:edm} and is fixed to $x^\ast$, respectively.}
  
  \vspace{0ex}
\end{figure}

Given the protocol $\varrho_{i}^\theta$ as in \eqref{eq:kld_rl_protocol}, we aim to design a payoff mechanism for the agents to asymptotically adopt the optimal strategy profile that minimizes a given cost $c(q)$. 
For instance, in Example~\ref{example:resource_collection}, if we design the payoff mechanism as $p = q$, the robots would select strategy~$i$ to take on task~$i$ and asymptotically minimize $\lim_{t \to \infty} \max_{1 \leq i \leq m} q_i(t)$, as discussed in \cite{9561809}. However, in many applications, such one-to-one correspondence between tasks and available strategies may not exist, and depending on the cost we want to minimize, such a simple payoff mechanism would not be the best design choice as we illustrate in Figure~\ref{fig:cost_example}. 

In addition, since the payoff mechanism depends on the vector $q(t)$, the mechanism would incentivize the agents to take on the tasks with larger $q_i(t)$. Hence, compared to other models that directly control the population state $x(t)$ to the optimal state $x^\ast$ (for instance, the model proposed in \cite{5161293}), our strategy revision model is more responsive to changes of $q(t)$ and hence reduces the task-associated variables $q(t)$ at a faster rate as we depict in Figure~\ref{fig:comparison_strategy_revision}.

Two examples of the cost function we consider are
\begin{itemize}
\item (square of) the 2-norm of $q$: $c(q) = \sum_{i=1}^m q_i^2$, and
\item the $\infty$-norm of $q$: $c(q) = \max_{1 \leq i \leq m} q_i$.
\end{itemize}
For the payoff mechanism design, we consider a linear model defined by a matrix $G \in \mathbb R^{n \times m}$ as follows:
\begin{align} \label{eq:incentive_mechanism}
  p = G q.
\end{align}
Our main problem investigates finding the matrix $G$ that allows the agents to asymptotically minimize the cost $c(q(t))$. We formally state the problem as follows.
\begin{problem} \label{problem:main}
  Given the dynamic model \eqref{eq:q_dynamics} of the task allocation game and the EDM \eqref{eq:edm}, compute the payoff matrix $G$ under which the cost $c(q(t))$ is asymptotically minimized.
\end{problem}

%%%%%%%%%%%%%%%%%%%%%%%%%%%%%%%%%%%%%%%%%%%%%%%%%%%%%%%%% 
\section{Payoff Matrix Design} \label{sec:incentive_design}

\begin{figure}[t!]
    \centering
    \includegraphics[trim={0in 0in 0in 0in}, clip, width=2.7in]{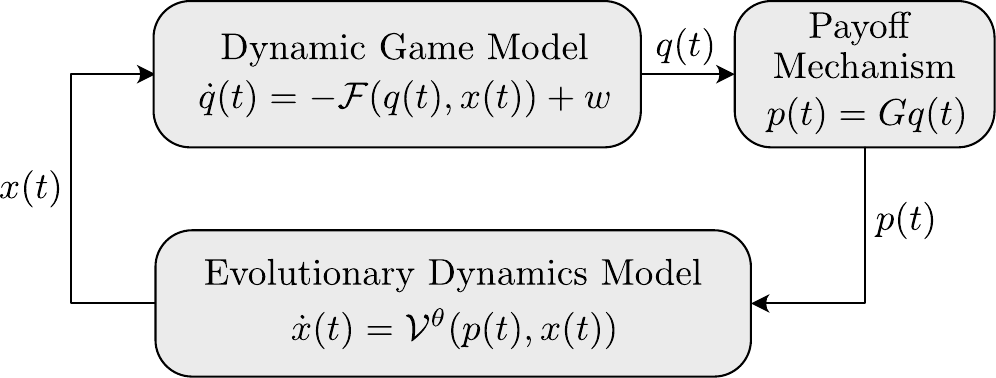}
    \caption{Feedback model in the task allocation game.}
    \label{fig:scheme}
    \vspace{-3ex}
\end{figure}

By interconnecting the dynamic model of the game \eqref{eq:q_dynamics}, the payoff mechanism \eqref{eq:incentive_mechanism}, and the EDM \eqref{eq:edm} with \eqref{eq:kld_rl_protocol} as its revision protocol, as illustrated in Figure~\ref{fig:scheme}, we can write the state equation of the resulting closed-loop model as follows:
\begin{subequations} \label{eq:feedback_model}
  \begin{align}
  &\begin{cases}
  \dot q(t) &= - \mathcal F(q(t), x(t)) + w \\
    p(t) &= G q(t) 
  \end{cases}
  \label{eq:feedback_model_a}\\
    &\dot x_i(t) = M \frac{\theta_i \exp(\eta^{-1} p_i(t))}{\sum_{l=1}^n \theta_l \exp(\eta^{-1} p_l(t))} - x_i(t). \label{eq:feedback_model_b}
  \end{align}
\end{subequations}
Given an initial condition $(q(0), \!x(0)) \!\in\! \mathbb R_+^m \!\times\! \mathbb X_M$, we assume the closed-loop model \eqref{eq:feedback_model} has a unique solution. Let $\mathbb S$ be the set of equilibrium states of \eqref{eq:feedback_model}.
The proper design of $G$ should ensure that the following two conditions hold.
\begin{enumerate}
  \renewcommand{\theenumi}{(R\arabic{enumi}}
\item \label{r1} The state $(q(t), x(t))$ converges to the stationary points of \eqref{eq:feedback_model_a}, i.e.,
it holds that
    $\lim_{t \to \infty} \inf_{(r,z) \in \mathbb O} (\|q(t) - r\|_2 + \|x(t) - z\|_2) = 0.$

\item \label{r2} When the closed-loop model \eqref{eq:feedback_model} reaches an equilibrium state $(q^\ast, x^\ast)$, it attains the minimum cost, i.e.,
    $c(q^\ast) = \inf_{(q, x) \in \mathbb O} c(q).$
\end{enumerate}

We adopt passivity tools \cite{Fox2013Population-Game, 9029756} to find technical conditions under which (R1) and (R2) are attained and use the conditions to design the payoff matrix $G$. The critical step in the convergence analysis (R1) is in establishing passivity for both \eqref{eq:feedback_model_a} and \eqref{eq:feedback_model_b} by finding a so-called $\delta$-storage function for \eqref{eq:feedback_model_a} and $\delta$-antistorage function for \eqref{eq:feedback_model_b}.\footnote{We refer to \cite[Definition~10]{9029756} and \cite[Definition~12]{9029756}, respectively, for the formal definitions of passivity for \eqref{eq:feedback_model_a} and \eqref{eq:feedback_model_b}.} Then, by constructing a Lyapunov function using the two storage functions, we establish convergence results for \eqref{eq:feedback_model}. 

To proceed, by \cite[Lemma~3]{park2023learning}, \eqref{eq:feedback_model_b} is $\delta$-passive and has the $\delta$-storage function $\mathcal S^\theta: \mathbb R^n \times \mathbb X_M \to \mathbb R_+$ given by
\begin{multline} \label{eq:delta_storage_function}
    \mathcal S^\theta (p, x) \!=\! \max_{z \in \mathbb X_M} ( p^T z \!-\! \eta \mathcal D(z || \theta) )
    \!-\! ( p^T x \!-\! \eta \mathcal D(x || \theta) ),
\end{multline}
where $\mathcal D ( \cdot \| \cdot )$ is the KL divergence.
Note that $\mathcal S^\theta$ satisfies 
\begin{subequations} \label{eq:delta_passivity_conditions}
\begin{align}
&\mathcal S^\theta (p, x) \!=\! 0 \Leftrightarrow \mathcal V^\theta(p,x) \!=\! 0 \Leftrightarrow \nabla_x^T \mathcal S^\theta(p,x) \mathcal V^\theta(p,x) \!=\! 0 \label{eq:delta_passivity_conditions_a} \\
&\mathcal S^\theta ( p(t), x(t) ) - \mathcal S^\theta \left( p(t_0), x(t_0) \right) \nonumber \\
  &\qquad\qquad\qquad \leq \int_{t_0}^t \dot p^T(\tau) \dot x(\tau) \, \mathrm d \tau , ~ \forall t \geq t_0 \geq 0 \label{eq:delta_passivity_conditions_b}
\end{align}
\end{subequations}
for any payoff vector trajectory $p(t), \, t \geq 0$. The mapping $\mathcal V^\theta = (\mathcal V_1^\theta, \cdots, \mathcal V_n^\theta)$ is the vector field of the EDM \eqref{eq:edm}.

The dynamic game model \eqref{eq:q_dynamics} is qualified as \textit{$\delta$-antipassive} \cite{9029756} if there is a \textit{$\delta$-antistorage function} $\mathcal L: \mathbb R_+^m \times \mathbb X_M \to \mathbb R_+$ satisfying the following two conditions:
\begin{subequations} \label{eq:delta_antipassivity_conditions}
\begin{align}
 \mathcal L ( q, x ) = 0 &\Leftrightarrow \mathcal F (q, x) = w  \nonumber \\
 &\Leftrightarrow \nabla_q^T \mathcal L(q,x) (\mathcal F(q,x) - w) = 0 \label{eq:delta_antipassivity_conditions_a}
 \end{align}
 \begin{align}
  &\mathcal L ( q(t), x(t) ) - \mathcal L \left( q(t_0), x(t_0) \right) \nonumber \\
  &\qquad\qquad \leq - \int_{t_0}^t \dot q^T(\tau) G^T \dot x(\tau) \, \mathrm d \tau , ~ \forall t \geq t_0 \geq 0, \label{eq:delta_antipassivity_conditions_b}
\end{align}
\end{subequations}
where \eqref{eq:delta_antipassivity_conditions_b} needs to hold for any given population state trajectory $x(t), ~ t \geq 0$. According to \eqref{eq:delta_antipassivity_conditions_a}, the function $\mathcal L(q,x)$ can be used to measure how far the state $(q,x)$ is from the equilibrium of \eqref{eq:feedback_model_a}. 
By their respective definitions \cite{9029756}, both $\mathcal S^\theta$ and $\mathcal L$ need to be continuously differentiable.

Recall $\mathbb O$ given as in \eqref{eq:q_dynamics_stationry_points}.
For $(q^\ast, x^\ast) \in \mathbb O$ satisfying
\begin{align} \label{eq:equilibrium_condition}
    x_i^\ast > 0 \implies p_i^\ast = \max_{1 \leq j \leq n} p_j^\ast, ~\forall i \in \{1, \cdots, n\}
\end{align}
with $p^\ast = G q^\ast$,
let us assign $\theta = x^\ast$ for \eqref{eq:feedback_model_b}. We can establish the following lemma.
\begin{lemma} \label{lemma:convergence_result}
   If 
   the dynamic game model \eqref{eq:feedback_model_a} is $\delta$-antipassive, then given that $q(t), \, t \geq 0$ is bounded, the state $(q(t), x(t))$ of the closed-loop model \eqref{eq:feedback_model} converges to $\mathbb S$. Also 
   $(q^\ast, x^\ast)$ is the equilibrium state of \eqref{eq:feedback_model} for all $\eta > 0$.
\end{lemma}

The proof of the lemma is given in Appendix.
Resorting to Lemma~\ref{lemma:convergence_result}, to meet the requirements (R1) and (R2), we need to construct the payoff matrix $G$ such a way that \eqref{eq:feedback_model_a} becomes $\delta$-antipassive and $(q^\ast, x^\ast) \in \mathbb O$ minimizing $c(q)$ is an equilibrium state of \eqref{eq:feedback_model}.
The following theorem states the technical conditions on $G$ that ensure (R1) and (R2). To state the theorem, we define a continuously differentiable mapping $g: \mathbb R_+^m \to \mathbb R_+^n$ that maps any $q \in \mathbb R_+^m$ to $y = g(q)$ satisfying $\mathcal F(q,y) = w$.\footnote{We remark that $g(q)$ does not necessarily belong to $\mathbb X_M$. We interpret $g(q)$ has the strategy profile that attains the equilibrium state for a given $q$ when there is no limit on the population mass $M$.} The statement of the theorem holds if such $g$ exists. 

\begin{theorem} \label{theorem:incentive_design}
  Let us define
  \begin{align}
    h_i(q,x) = (\mathcal F_i(q_i,x) - w_i) \, (x - g (q)), ~ i \in \{1, \cdots, m\} \nonumber
  \end{align}
  and let $(q^\ast, x^\ast)$ be the stationary point of \eqref{eq:feedback_model_a} attaining the minimum cost $\inf_{(q,x) \in \mathbb O} c(q)$. Suppose the matrix $G$ satisfies 
  \vspace{-1em}
\begin{subequations} \label{eq:G_requirement}
  \begin{align} 
  &G \nabla_x \mathcal F(q,x) = \nabla_x^T \mathcal F(q,x) G^T, \,\forall (q,x) \in \mathbb R_+^m \times \mathbb R_+^n \label{eq:GF_symmetric_requirement} \\
  &h_i^T(q,x) G_i^{\tiny \text{col}} > 0, ~ \forall (q,x) \notin \mathbb O, \,\forall i \in \{1, \cdots, m\} \label{eq:G_column_requirement} \\
  &(G_i^{\tiny \text{row}} - G_j^{\tiny \text{row}}) x_i^\ast q^\ast \geq 0, ~ \forall i,j \in \{1, \cdots, n\}, \label{eq:G_row_requirement}
  \end{align}
\end{subequations}
where $G_i^{\tiny \text{col}}$ and $G_i^{\tiny \text{row}}$ are the column and row vectors of $G$ defined as in \eqref{eq:G_representation}, respectively.
The dynamic game model \eqref{eq:feedback_model_a} is $\delta$-antipassive and $(q^\ast, x^\ast)$ is an equilibrium state of \eqref{eq:feedback_model} with $\theta = x^\ast$ for any $\eta > 0$. 
\end{theorem}

The proof of the theorem is given in Appendix.
Under the condition \eqref{eq:G_column_requirement}, whenever $q_i(t)$ is increasing, i.e., $\dot q_i(t) = -\mathcal F_i(q_i(t), x(t)) + w_i > 0$, the matrix $G$ incentivizes the agents to revise their strategies toward $g(q)$, which is the strategy profile required to make the rate $\dot q(t)$ to zero. In other words, $G$ is designed to encourage the agents to select strategies that reduce the rate $\dot q(t)$.

\begin{proposition} \label{proposition:unique_equilibrium}
Let $(q^\ast, x^\ast)$ be the stationary point of \eqref{eq:feedback_model_a} attaining the minimum cost $\inf_{(q,x) \in \mathbb O} c(q)$. Consider the closed-loop model \eqref{eq:feedback_model} for which $\theta = x^\ast$ and the payoff matrix $G$ satisfies \eqref{eq:G_requirement}. As the parameter $\eta$ of \eqref{eq:feedback_model_b} increases, $(q^\ast, x^\ast)$ becomes the unique equilibrium state of \eqref{eq:feedback_model}. In other words, it holds that
    $\lim_{\eta \to \infty} \sup_{(\bar q, \bar x) \in \mathbb S} \mathcal D(\bar x \,\|\, x^\ast) = 0$,
where $\mathbb S$ is the set of equilibrium states of \eqref{eq:feedback_model}. 
\end{proposition}

The proof of the proposition is provided in Appendix.
In conjunction with Lemma~\ref{lemma:convergence_result} and Theorem~\ref{theorem:incentive_design}, Proposition~\ref{proposition:unique_equilibrium} implies that as $\eta$ becomes sufficiently large, the state trajectory $(q(t), x(t)),~ t \geq 0$ converges to near the optimal state $(q^\ast, x^\ast)$. According to \eqref{eq:kld_rl_protocol}, we note that smaller $\eta$ is desired to make the agent strategy revision responsive to changes in $p(t)$ and also in $q(t)$. Hence, a good practice is to use smaller $\eta$ at the beginning of the task allocation game, and if needed, as $\dot q(t)$ goes to zero, the agents can gradually increase the value of $\eta$ to ensure that 
$x(t)$ converges to $x^\ast$.

%%%%%%%%%%%%%%%%%%%%%%%%%%%%%%%%%%%%%%%%%%%%%%%%%%%%%%%%% 
\section{Simulations} \label{sec:simulations}
We use Examples~\ref{example:resource_collection} and \ref{example:heterogeneous_sensing} to illustrate our main results and discuss how the cost function and parameters of the dynamic model \eqref{eq:q_dynamics} affect the payoff matrix design. In both examples, we select the following fixed parameters $M=1$, $R_i = 3.5$, $\alpha_i = 0.05$, and $\beta_i = 1$ for \eqref{eq:q_dynamics_example_1} and \eqref{eq:q_dynamics_example_2},
and $\eta = 0.001$ for \eqref{eq:feedback_model_b}.\footnote{We select $\eta = 0.001$ as all population state trajectories in the simulations converge to the optimal $x^\ast$ with the small positive $\eta$.}
We use two different cost functions $c(q)$ for Example~\ref{example:resource_collection} and two distinct growth rates $w$ for Example~\ref{example:heterogeneous_sensing}.

\subsection{Computation of $G$} \label{sec:G_computation}
We explain the steps to compute $G$. First, note that \eqref{eq:GF_symmetric_requirement} is satisfied if $G$ has the following structures:
\begin{enumerate}
    \item For Example~\ref{example:resource_collection}, $G_{ij} = 0$ if $i \neq j$.
    
    \item For Example~\ref{example:heterogeneous_sensing}, $G_{ij} = 0$ if $i \notin \mathbb N_j$ and $G_{ij} = G_j$ otherwise, where $G_j$ is a real number.
\end{enumerate}

Then, we find $(q^\ast, x^\ast) \in \mathbb O$ that minimizes the cost function $c(q)$ using the following optimization.
\begin{align}
  \min_{(q, x) \in \mathbb R_+^m \times \mathbb X_M} c(q) \text{ subject to } \mathcal F(q,x) = w.
\end{align}
Note that since $\mathcal F$ is a nonlinear mapping, the optimization can be non-convex and the solution we find is locally optimal.

Once we find $(q^\ast, x^\ast)$, we compute the matrix $G$ satisfying \eqref{eq:G_requirement} for which we first need to find the mapping $g$. Instead of explicitly finding $g$, we draw random samples $\{(q_s, x_s)\}_{s=1}^S \subset \mathbb R_+^m \times \mathbb X_M$ and find $y_s \in \mathbb R_+^n$ that minimizes $\| \mathcal F(q_s, y_s) - w \|_2^2$ for each sample $(q_s, x_s)$. Note that assuming $\nabla_x \mathcal F(q,x)$ has full rank at $(q_s, y_s)$, which is the case in both examples, the minimizer $y_s$ satisfies $\mathcal F(q_s, y_s) = w$. 

As the last step, the design of $G$ can be formulated as the following linear programming:
\begin{align} \label{eq:computing_G}
  &\min_{G \in \mathbb R^{n \times m}} 1 \\
  &\begin{aligned}
    \text{subject to } & (\mathcal F_i(q_{s,i},x_s) - w_i) \, (x_s - y_s)^T G_i^{\tiny \text{col}} > 0, \nonumber \\
    &\qquad\qquad ~ \forall i \in \{1, \cdots, m\}, ~ \forall s \in \{1, \cdots, S \} \nonumber \\
    &(G_i^{\tiny \text{row}} - G_j^{\tiny \text{row}}) x_i^\ast q^\ast \geq 0, ~ \forall i,j \in \{1, \cdots, n\}, \nonumber
  \end{aligned}
\end{align}
where $q_{s,i}$ is the $i$-th element of $q_s = (q_{s,1}, \cdots, q_{s,m})$.
Since we evaluate the condition \eqref{eq:G_column_requirement} using a finite number of sampled points $\{(q_s, x_s)\}_{s=1}^S$, we would obtain an approximate solution satisfying \eqref{eq:G_requirement} only at the sampled points. However, as the sample size $S$ tends to infinity, the solution $G$ is more likely to satisfy \eqref{eq:G_requirement} over the entire state space $\mathbb R_+^m \times \mathbb X_M$.

\subsection{Simulation results for Example~\ref{example:resource_collection} ($m=4, n=4$)} \label{sec:simulation_1}
\begin{figure}
  \center
  \subfigure[Population state $x(t)$]{
    \includegraphics[trim={.1in .1in .1in 0in}, clip, width=1.40in]{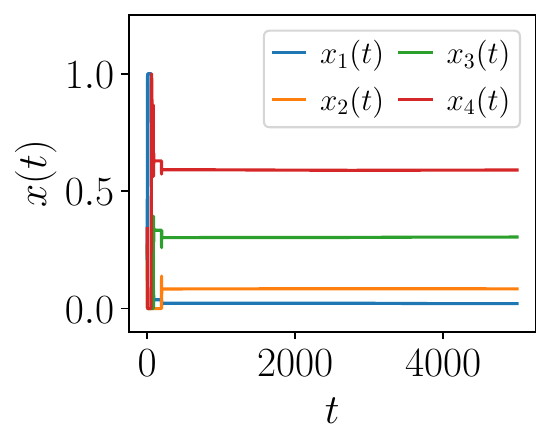}
    \label{fig:simulation_01_a}
  }
  \subfigure[Task-associated vector $q(t)$]{
    \includegraphics[trim={.1in .1in .1in .0in}, clip, width=1.40in]{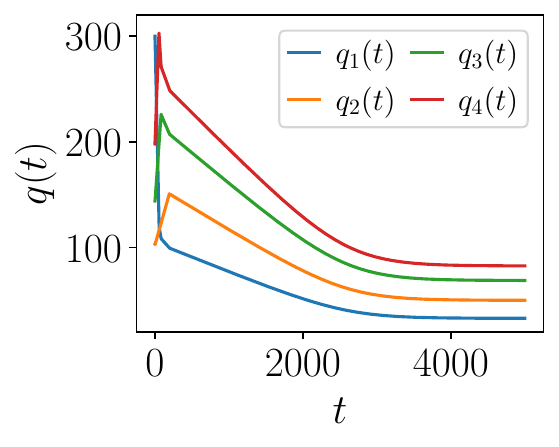}
    \label{fig:simulation_01_b}
  }
  
  \caption{Graphs depicting the trajectories of the (a) population state $x(t)$ and (b) task-associated vector $q(t)$ derived by the closed-loop model \eqref{eq:feedback_model} in Example~\ref{example:resource_collection} using the cost function $c(q) = \sum_{i=1}^4 q_i^2$.}
  \label{fig:simulation_01}
  \center
  \subfigure[Population state $x(t)$]{
    \includegraphics[trim={.1in .1in .1in 0in}, clip, width=1.40in]{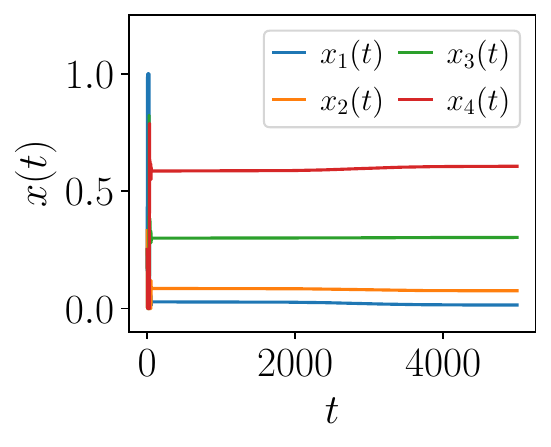}
    \label{fig:simulation_02_a}
  }
  \subfigure[Task-associated vector $q(t)$]{
    \includegraphics[trim={.1in .1in .1in .0in}, clip, width=1.40in]{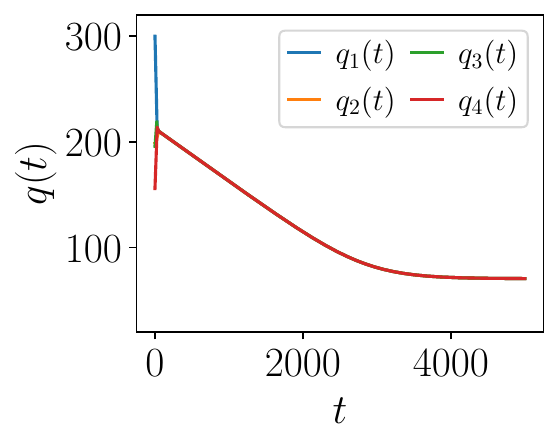}
    \label{fig:simulation_02_b}
  }
  
  \caption{Graphs depicting the trajectories of the (a) population state $x(t)$ and (b) task-associated vector $q(t)$ derived by the closed-loop model \eqref{eq:feedback_model} in Example~\ref{example:resource_collection} using the cost function $c(q) = \max_{1 \leq i \leq 4} q_i$.}
  \label{fig:simulation_02}
  \vspace{-3ex}
\end{figure}

Using the methods explained in Section~\ref{sec:G_computation}, we compute the optimal state $(q^\ast, x^\ast)$ minimizing \textbf{i)} $c(q) = \sum_{i=1}^m q_i^2$ and \textbf{ii)} $c(q) = \max_{1 \leq i \leq m} q_i$, where we use the fixed growth rate $w = (0.05, 0.25, 1.00, 2.00)$ for both cases. Then, we design the payoff matrix $G$ using \eqref{eq:computing_G} as follows.
\begin{enumerate}
\renewcommand{\theenumi}{\roman{enumi}}

\item For $c(q) = \sum_{i=1}^m q_i^2$,
\begin{align}
\footnotesize
    G = \begin{pmatrix} 
    1.00 & 0.00 & 0.00 & 0.00 \\
    0.00 & 0.66 & 0.00 & 0.00 \\
    0.00 & 0.00 & 0.48 & 0.00 \\
    0.00 & 0.00 & 0.00 & 0.40
    \end{pmatrix}. \nonumber
\end{align}

\item For $c(q) = \max_{1 \leq i \leq m} q_i$,
\begin{align}
\footnotesize
    G = \begin{pmatrix} 
    1.00 & 0.00 & 0.00 & 0.00 \\
    0.00 & 1.00 & 0.00 & 0.00 \\
    0.00 & 0.00 & 1.00 & 0.00 \\
    0.00 & 0.00 & 0.00 & 1.00
    \end{pmatrix}. \nonumber
\end{align}

\end{enumerate}

Note that when we use the $\infty$-norm to define the cost $c(q)$, the optimal design of $G$ equally incentivizes the agents proportional to the remaining jobs $q$.
On the other hand, when the $2$-norm is used, given that the values of $q_1(t), \cdots, q_4(t)$ are equal, the 
payoff matrix 
$G$ assigns the highest payoff to strategy~$1$ and the lowest payoff to strategy~$4$. Recall that under the pre-selected growth rate $w$, task~$1$ has the lowest growth rate and task~$4$ has the highest, and hence maintaining lower $q_1(t)$ is easier -- it needs a less number of agents -- than $q_4(t)$. Hence, under the $2$-norm cost function, the agents prioritize to carry out
the tasks with lower growth rates. 

Figures~\ref{fig:simulation_01} and \ref{fig:simulation_02} depict the resulting trajectories for $x(t)$ and $q(t)$. Notice that the population states at the equilibrium in the two cases are similar; however, the  trajectories for $q(t)$ are different and, hence, so do the costs evaluated along the trajectories as we discussed in Figure~\ref{fig:cost_example}. We observe that there is a large variation in the agent strategy revision at the beginning of the simulations as the agents repeatedly switch among the strategies to reduce $q_i(t)$ with a larger value.
  
\subsection{Simulation results for Example~\ref{example:heterogeneous_sensing} ($m=4, n=6$)} \label{sec:simulation_2}
\begin{figure}
  \center
  \subfigure[Population state $x(t)$]{
    \includegraphics[trim={.1in .1in .1in 0in}, clip, width=1.45in]{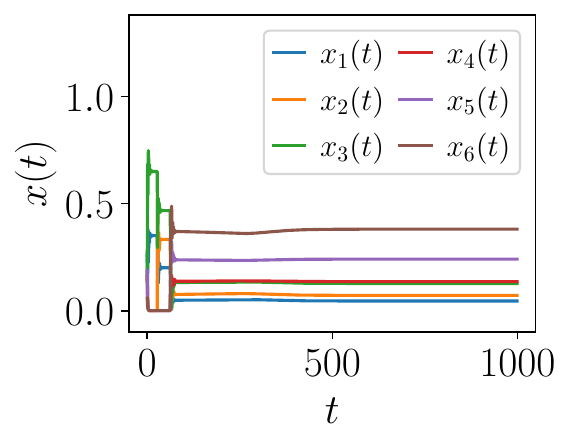}
    \label{fig:simulation_03_a}
  }
  \subfigure[Task-associated vector $q(t)$]{
    \includegraphics[trim={.1in .1in .1in .0in}, clip, width=1.45in]{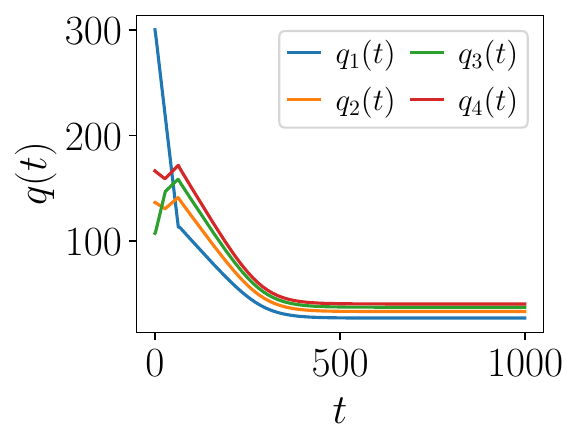}
    \label{fig:simulation_03_b}
  }
  
  \caption{Graphs depicting the trajectories of the (a) population state $x(t)$ and (b) task-associated vector $q(t)$ derived by the closed-loop model \eqref{eq:feedback_model} in Example~\ref{example:heterogeneous_sensing} using the growth rate $w = (0.5, 1.0, 1.5, 2.0)$.}
  \label{fig:simulation_03}

  \center
  \subfigure[Population state $x(t)$]{
    \includegraphics[trim={.1in .1in .1in 0in}, clip, width=1.45in]{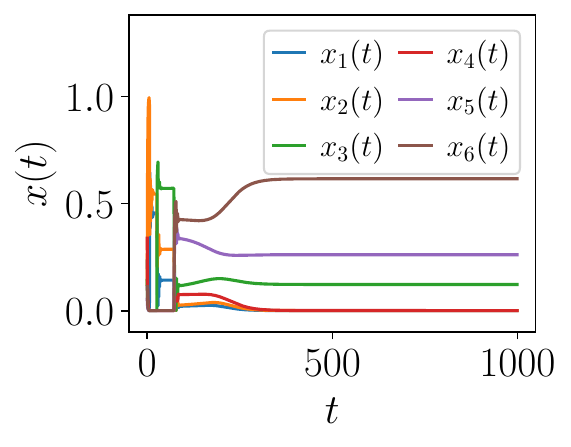}
    \label{fig:simulation_04_a}
  }
  \subfigure[Task-associated vector $q(t)$]{
    \includegraphics[trim={.1in .1in .1in .0in}, clip, width=1.45in]{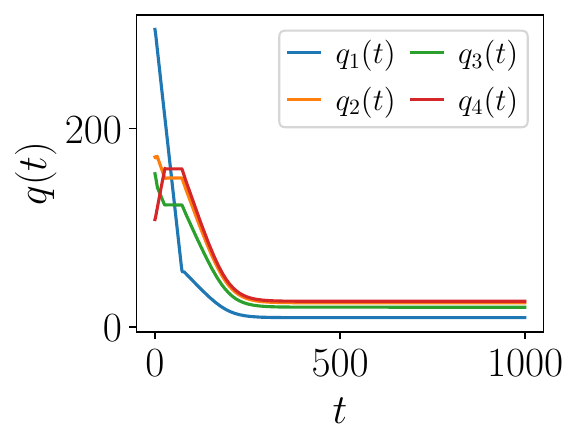}
    \label{fig:simulation_04_b}
  }
  
  \caption{Graphs depicting the trajectories of the (a) population state $x(t)$ and (b) task-associated vector $q(t)$ derived by the closed-loop model \eqref{eq:feedback_model} in Example~\ref{example:heterogeneous_sensing} using the growth rate $w = (0.1, 0.5, 1.0, 2.0)$.}
  \label{fig:simulation_04}
  \vspace{-3ex}
\end{figure}

We consider that there are 4 target states and 4 types of sensors each of which can measure a single state of the target. Each mobile unit can be equipped with two types of sensors and we define a strategy based on a pair of sensors employed on a mobile unit. According to the strategy definition, we can define the set $\mathbb N_i$ in \eqref{eq:q_dynamics_example_2} as the strategies that can measure the $i$-th target state: $\mathbb N_1 = \{1, 2, 3\}$, $\mathbb N_2 = \{1, 4, 5\}$, $\mathbb N_3 = \{2, 4, 6\}$, and $\mathbb N_4 = \{3, 5, 6\}$. We use the square of the $2$-norm to define the cost function, i.e., $c(q) = \sum_{i=1}^m q_i^2$.

We 
design 
$G$ with two distinct growth rates as follows.
\begin{enumerate}
    \renewcommand{\theenumi}{\roman{enumi}}

    \item For $w = (0.5, 1.0, 1.5, 2.0)$,
    \begin{align}
    \footnotesize
    G = \begin{pmatrix} 
    1.00 & 0.81 & 0.00 & 0.00 \\
    1.00 & 0.00 & 0.72 & 0.00 \\
    1.00 & 0.00 & 0.00 & 0.67 \\
    0.00 & 0.81 & 0.72 & 0.00 \\
    0.00 & 0.81 & 0.00 & 0.67 \\
    0.00 & 0.00 & 0.72 & 0.67
    \end{pmatrix}. \nonumber
    \end{align}

    \item For $w = (0.1, 0.5, 1.0, 2.0)$,
    \begin{align}
    \footnotesize
    G = \begin{pmatrix} 
    1.68 & 0.99 & 0.00 & 0.00 \\
    1.68 & 0.00 & 0.80 & 0.00 \\
    1.68 & 0.00 & 0.00 & 0.67 \\
    0.00 & 0.99 & 0.80 & 0.00 \\
    0.00 & 0.99 & 0.00 & 0.67 \\
    0.00 & 0.00 & 0.80 & 0.67
    \end{pmatrix}. \nonumber
\end{align}
\end{enumerate}

By comparing the above two payoff matrices, we can infer that the optimal $G$ assigns higher payoffs to the strategies as their respective growth rates become smaller. Figures~\ref{fig:simulation_03} and \ref{fig:simulation_04} depict the resulting trajectories for $x(t)$ and $q(t)$. Notably, as the growth rate of the $4$-th target state becomes relatively higher than those of other target states, more agents adopt strategies~$3,5$, and $6$, which can be used to measure the $4$-th state. Similar to the simulation results in Section~\ref{sec:simulation_1}, we can observe a large variation in the agent strategy revision at the beginning of the simulations as the agents responsively revise their strategies based on the value of $q_i(t)$.

%%%%%%%%%%%%%%%%%%%%%%%%%%%%%%%%%%%%%%%%%%%%%%%%%%%%%%%%% 
\section{Conclusions} \label{sec:conclusions}
We investigated the design of the payoff mechanism in the task allocation games. The mechanism determines payoffs $p$ given the vector $q$ that quantifies the amount of jobs in the assigned tasks to the agents, and the payoffs incentivize the agents to repeatedly revise their strategy selection.
We discussed how to design the payoff matrix $G$ using the passivity tools to ensure that the agents asymptotically attain the optimal strategy profile.
Using the numerical examples, we demonstrated how our results can be used to design $G$ and how the parameters of the dynamic game model affect the optimal design of $G$. 
For future directions, we plan to consider the design of nonlinear payoff mechanisms $p = G(q)$, and to explore the idea of learning the dynamic model and computing $G$ alongside the model learning.

%%%%%%%%%%%%%%%%%%%%%%%%%%%%%%%%%%%%%%%%%%%%%%%%%%%%%%%%%
% \newpage
\appendix
\subsection{Proof of Lemma~\ref{lemma:convergence_result}}
From \eqref{eq:delta_passivity_conditions_b} and \eqref{eq:delta_antipassivity_conditions_b}, we can derive 
\begin{align}
    &\frac{\mathrm d}{\mathrm dt} \left( \mathcal S^\theta(p(t), x(t)) + \mathcal L (q(t), x(t))\right) \nonumber \\
    &= \nabla_x^T \mathcal S^\theta(p(t), x(t)) \mathcal V^\theta (p(t), x(t)) \nonumber \\
    &\qquad + \nabla_q^T \mathcal L(q(t), x(t)) (-\mathcal F(q(t), x(t)) + w) \leq 0,
\end{align}
where by \eqref{eq:delta_passivity_conditions_a} and \eqref{eq:delta_antipassivity_conditions_a}, the equality holds if and only if $\mathcal S^\theta(p(t), x(t)) = \mathcal L(q(t), x(t)) = 0$. Therefore, by LaSalle's invariance principle \cite{Khalil:1173048}, if $q(t)$ is bounded, then the trajectory $(q(t), x(t)), \, t \geq 0$ converges to the equilibrium states of \eqref{eq:feedback_model}.

By the definition of the KLD-RL protocol \cite{9483414, park2023learning}, with $\theta = x^\ast$, $(q^\ast, x^\ast)$ is an equilibrium state of \eqref{eq:feedback_model} for any $\eta > 0$ if it holds that 
    $x^\ast \in \argmax_{z \in \mathbb X_M} (z^T Gq^\ast)$,
which can be validated by \eqref{eq:equilibrium_condition}.
\qed

\subsection{Proof of Theorem~\ref{theorem:incentive_design}}
We define a $\delta$-antistorage function $\mathcal L: \mathbb R_+^m \times \mathbb X_M \to \mathbb R_+$ using a line integral along a curve $s$
from $g(q)$ to $x$:
\begin{align}
  \mathcal L(q, x) = \int_{g(q)}^x G (\mathcal F (q,s) -w )\bullet \mathrm ds.
\end{align}
Recall that $g(q) \in \mathbb R_+^n$ is such that $\mathcal F(q,g(q)) = w$.
By evaluating the line integral along the curve given by $s(\tau) = \tau x + (1-\tau) g(q), ~ \tau \in [0, 1]$, we can derive
\begin{align} \label{eq:antistorage_function_01}
  \mathcal L(q, x) = (x - g(q))^T G \int_{0}^1 \left( \mathcal F (q,s (\tau)) - w \right) \mathrm d\tau.
\end{align}
Note that for every $\tau$ in $(0, 1)$, it holds that
\begin{align} \label{eq:antistorage_function_02}
  &(x - g(q))^T G \left( \mathcal F (q,s(\tau)) - w \right) \nonumber \\
  &=\frac{1}{\tau}\sum_{i=1}^m (s(\tau) - g(q))^T G_i^{\tiny \text{col}} \left( \mathcal F_i (q_i, s(\tau)) - w_i \right).
\end{align}
Hence, under \eqref{eq:G_column_requirement}, we can verify that $\mathcal L(q,x) \geq 0$ where the equality holds if and only if $\mathcal F (q,x) = w$.
In what follows, we show that $\mathcal L$ satisfies \eqref{eq:delta_antipassivity_conditions_a} and \eqref{eq:delta_antipassivity_conditions_b}.
To begin with, using \eqref{eq:GF_symmetric_requirement}, we can establish
\begin{align}
    \nabla_x \mathcal L(q,x) 
    &= G \int_{0}^1 \left( \mathcal F (q,s (\tau)) - w \right) \mathrm d\tau \nonumber \\
    &\quad + \int_{0}^1 \nabla_z^T \mathcal F (q, z) \Big|_{z = s (\tau)} \tau \, \mathrm d\tau \, G^T (x - g(q)) \nonumber \\
    &= G \int_{0}^1 \frac{\mathrm d}{\mathrm d \tau} \tau \mathcal F (q,s (\tau)) \, \mathrm d\tau - G w \nonumber \\
    &= G \left( \mathcal F (q,x) - w \right).
\end{align}
Hence, we can derive
\begin{align}
  \nabla_x^T \mathcal L (q,x) \dot x = (\mathcal F (q,x) - w)^T G^T \dot x = -\dot p^T \dot x.
\end{align}
Similarly, the partial derivative of $\mathcal L (q,x)$ with respect to $q$ can be computed as
\begin{align}
    &\nabla_q \mathcal L (q,x) \nonumber \\
    &= -\nabla_q^T g(q) \int_{0}^1 G \left( \mathcal F (q,s (\tau)) - w \right) \mathrm d\tau \nonumber \\
    &\quad +  \int_{0}^1 \bigg( \nabla_q^T G \mathcal F (q,s (\tau)) \nonumber \\ 
    &\qquad + \nabla_q^T g(q) \nabla_z^T G \mathcal F (q, z) \Big|_{z = s(\tau)} (1-\tau) \bigg) \mathrm d\tau \, (x - g(q)) \nonumber \\
    &=\int_{0}^1 \nabla_q^T \mathcal F (q,s (\tau)) \, \mathrm d\tau \, G^T (x - g(q)) \nonumber \\
    &\quad +  \nabla_q^T g(q) \underbrace{\int_{0}^1 G \bigg( \frac{\mathrm d}{\mathrm d \tau} \mathcal F (q, s(\tau)) (1-\tau) + w\bigg) \mathrm d\tau}_{=0}.
\end{align}
Therefore, we can derive
\begin{align}
  &\nabla_q^T \mathcal L(q,x) \dot q \nonumber \\
  &= (x - g(q))^T \int_{0}^1 \nabla_q G \mathcal F (q,s (\tau)) \, \mathrm d\tau \, (-\mathcal F(q,x) + w). \nonumber
\end{align}
Then, the time derivative of $\mathcal L$ becomes
\begin{multline}
  \frac{\mathrm d}{\mathrm dt} \mathcal L ( q, x ) = (x - g(q))^T \\ \times \int_{0}^1 \nabla_q G \mathcal F (q,s (\tau)) \, \mathrm d\tau \, (-\mathcal F(q,x) + w) - \dot p^T \dot x
\end{multline}
Hence, for $\mathcal L$ to satisfy \eqref{eq:delta_antipassivity_conditions_b}, it suffices to show that the following inequality holds.
\begin{align} \label{eq:delta_antistorage_function_negativity}
  (x\! -\! g(q))^T\! \int_{0}^1 \nabla_q G \mathcal F (q,s (\tau)) \mathrm d\tau (-\mathcal F(q,x) \!+\! w) \!\leq\! 0.
\end{align}

By Assumption~\ref{assumption:on_F}, we can rewrite the above equation as
\begin{align}
  &(x - g(q))^T \int_{0}^1 \nabla_q G \mathcal F (q,s (\tau)) \, \mathrm d\tau \, (-\mathcal F(q,x) + w) \nonumber \\
  &=\!\sum_{i=1}^m \!\int_{0}^1 \!\frac{\partial \mathcal F_i}{\partial q_i} (q_i,s (\tau)) \mathrm d\tau 
  (x \!-\! g(q))^T G_i^{\tiny \text{col}} (-\mathcal F_i(q_i,x) \!+\! w_i), \nonumber
\end{align}
where $G_i^{\tiny \text{col}}$ is the $i$-th column vector of $G$ defined as in \eqref{eq:G_representation}.
Consequently, by Assumption~\ref{assumption:on_F}, the condition \eqref{eq:G_column_requirement} ensures  \eqref{eq:delta_antistorage_function_negativity}, where the equality holds if and only if $(q,x) \in \mathbb O$. Hence, in conjunction with the fact that $\mathcal L(q,x) = 0 \Leftrightarrow \mathcal F(q,x) = w$, we can validate that \eqref{eq:delta_antipassivity_conditions_a} holds.

Recall that, according to \eqref{eq:equilibrium_state}, for $(q^\ast, x^\ast)$ minimizing the cost function $c(q)$ to be the equilibrium state of the closed-loop model \eqref{eq:feedback_model}, it suffices to show
\begin{align}
  x_i^\ast > 0
  & \implies p_i^\ast = \max_{1 \leq j \leq n} p_j^\ast \nonumber \\
  &\implies G_i^{\tiny \text{row}} q^\ast = \max_{1 \leq j \leq n} G_j^{\tiny \text{row}} q^\ast \nonumber \\
  &\implies (G_i^{\tiny \text{row}} - G_j^{\tiny \text{row}}) q^\ast \geq 0
\end{align}
holds for all $i,j$ in $\{1, \cdots, n\}$, where $G_i^{\tiny \text{row}}$ is the $i$-th row vector of $G$ defined as in \eqref{eq:G_representation}. Hence, the condition \eqref{eq:G_row_requirement} ensures that $(q^\ast, x^\ast)$ is the equilibrium state of \eqref{eq:feedback_model}. This completes the proof.  \qed

\subsection{Proof of Proposition~\ref{proposition:unique_equilibrium}}
First of all, according to Lemma~\ref{lemma:convergence_result} and \eqref{eq:G_row_requirement}, $(q^\ast, x^\ast)$ is an equilibrium point of \eqref{eq:feedback_model}.
By the definition of the KLD-RL protocol \cite{9483414, park2023learning}, with $\theta = x^\ast$, for any other equilibrium state $(\bar q, \bar x)$ of \eqref{eq:feedback_model}, it holds that
\begin{align} \label{eq:kld_rl_equilibrium_condition}
    (\bar x - x^\ast)^T G \bar q  \geq \eta \mathcal D(\bar x \,\|\, x^\ast).
\end{align}
We prove the statement of the proposition by showing that for any fixed positive constant $\epsilon$, when $\eta$ is sufficiently large, there is no equilibrium state $(\bar q, \bar x)$ of \eqref{eq:feedback_model} satisfying $\mathcal D(\bar x \,\|\, x^\ast) \geq \epsilon$.

By contradiction, for each $\eta > 0$, suppose there is an equilibrium state $(\bar q, \bar x)$ for which $\mathcal D(\bar x \,\|\, x^\ast) \geq \epsilon$ holds.
When $\eta$ is sufficiently large, for $(\bar q, \bar x)$ to be an equilibrium state of \eqref{eq:feedback_model}, $(\bar x - x^\ast)^T G \bar q$ needs to be large enough to satisfy \eqref{eq:kld_rl_equilibrium_condition}. 
Note that
    $(\bar x - x^\ast)^T G \bar q = \sum_{i=1}^m (\bar x - x^\ast)^T G_i^{\text{col}} \bar q_i$.
Let $i$ be an index 
for which $(\bar x - x^\ast)^T G_i^{\text{col}} \bar q_i$ becomes arbitrarily large and so does $\bar q_i$ 
as $\eta$ increases.
According to \eqref{eq:G_column_requirement}, when $q = \bar q$ and $x = x^\ast$, it holds that $g(\bar q) = \bar x$ and hence we have
\begin{align} \label{eq:G_design_requirement}
    (\mathcal F_i(\bar q_i, x^\ast) - w_i) (x^\ast - \bar x)^T G_i^{\text{col}} > 0.
\end{align}
By Assumption~\ref{assumption:on_F} and by the fact that $(q^\ast, x^\ast)$ is an equilibrium state of \eqref{eq:feedback_model}, as $\bar q_i$ becomes arbitrarily large, $\mathcal F_i(\bar q_i, x^\ast) > w_i$ holds in which case by \eqref{eq:G_design_requirement}, it holds that $(\bar x - x^\ast)^T G_i^{\text{col}} < 0$. However, this contradicts the requirement that $(\bar x - x^\ast)^T G_i^{\text{col}} \bar q_i$ takes a large positive value. This completes the proof.
\qed

%%%%%%%%%%%%%%%%%%%%%%%%%%%%%%%%%%%%%%%%%%%%%%%%%%%%%%%%% 
\balance
\bibliographystyle{IEEEtran}
\bibliography{IEEEabrv,references}

\begin{thebibliography}{10}
\providecommand{\url}[1]{#1}
\csname url@rmstyle\endcsname
\providecommand{\newblock}{\relax}
\providecommand{\bibinfo}[2]{#2}
\providecommand\BIBentrySTDinterwordspacing{\spaceskip=0pt\relax}
\providecommand\BIBentryALTinterwordstretchfactor{4}
\providecommand\BIBentryALTinterwordspacing{\spaceskip=\fontdimen2\font plus
\BIBentryALTinterwordstretchfactor\fontdimen3\font minus
  \fontdimen4\font\relax}
\providecommand\BIBforeignlanguage[2]{{%
\expandafter\ifx\csname l@#1\endcsname\relax
\typeout{** WARNING: IEEEtran.bst: No hyphenation pattern has been}%
\typeout{** loaded for the language `#1'. Using the pattern for}%
\typeout{** the default language instead.}%
\else
\language=\csname l@#1\endcsname
\fi
#2}}

\bibitem{9561809}
S.~Park, Y.~D. Zhong, and N.~E. Leonard, ``Multi-robot task allocation games in
  dynamically changing environments,'' in \emph{2021 IEEE International
  Conference on Robotics and Automation (ICRA)}, 2021.

\bibitem{681242}
L.~Parker, ``Alliance: an architecture for fault tolerant multirobot
  cooperation,'' \emph{IEEE Transactions on Robotics and Automation}, vol.~14,
  no.~2, pp. 220--240, 1998.

\bibitem{6558009}
C.~Yang, L.~Kaplan, E.~Blasch, and M.~Bakich, ``Optimal placement of
  heterogeneous sensors for targets with {Gaussian} priors,'' \emph{IEEE
  Transactions on Aerospace and Electronic Systems}, vol.~49, no.~3, pp.
  1637--1653, 2013.

\bibitem{1677945}
L.~Parker and F.~Tang, ``Building multirobot coalitions through automated task
  solution synthesis,'' \emph{Proceedings of the IEEE}, vol.~94, no.~7, pp.
  1289--1305, 2006.

\bibitem{8439076}
I.~Jang, H.-S. Shin, and A.~Tsourdos, ``Anonymous hedonic game for task
  allocation in a large-scale multiple agent system,'' \emph{IEEE Transactions
  on Robotics}, vol.~34, no.~6, pp. 1534--1548, 2018.

\bibitem{5674046}
W.~Saad, Z.~Han, T.~Basar, M.~Debbah, and A.~Hjorungnes, ``Hedonic coalition
  formation for distributed task allocation among wireless agents,'' \emph{IEEE
  Transactions on Mobile Computing}, vol.~10, no.~9, 2011.

\bibitem{5072249}
H.-L. Choi, L.~Brunet, and J.~P. How, ``Consensus-based decentralized auctions
  for robust task allocation,'' \emph{IEEE Transactions on Robotics}, vol.~25,
  no.~4, pp. 912--926, 2009.

\bibitem{10.1007/978-3-030-20915-5_54}
S.~Amaya and A.~Mateus, ``Tasks allocation for rescue robotics: A replicator
  dynamics approach,'' in \emph{Artificial Intelligence and Soft Computing},
  L.~Rutkowski, R.~Scherer, M.~Korytkowski, W.~Pedrycz, R.~Tadeusiewicz, and
  J.~M. Zurada, Eds.\hskip 1em plus 0.5em minus 0.4em\relax Cham: Springer
  International Publishing, 2019, pp. 609--621.

\bibitem{5161293}
S.~Berman, A.~Halasz, M.~A. Hsieh, and V.~Kumar, ``Optimized stochastic
  policies for task allocation in swarms of robots,'' \emph{IEEE Transactions
  on Robotics}, vol.~25, no.~4, pp. 927--937, 2009.

\bibitem{doi:10.1080/00207179.2016.1231422}
A.~Pashaie, L.~Pavel, and C.~J. Damaren, ``A population game approach for
  dynamic resource allocation problems,'' \emph{International Journal of
  Control}, vol.~90, no.~9, pp. 1957--1972, 2017.

\bibitem{MARDEN20123075}
J.~R. Marden, ``State based potential games,'' \emph{Automatica}, vol.~48,
  no.~12, pp. 3075--3088, 2012.

\bibitem{6459524}
N.~Li and J.~R. Marden, ``Designing games for distributed optimization,''
  \emph{IEEE Journal of Selected Topics in Signal Processing}, vol.~7, no.~2,
  pp. 230--242, 2013.

\bibitem{Sandholm2010-SANPGA-2}
W.~H. Sandholm, \emph{Population Games and Evolutionary Dynamics}.\hskip 1em
  plus 0.5em minus 0.4em\relax MIT Press, 2011.

\bibitem{EDM_water_passivity}
E.~Ramirez-Llanos and N.~Quijano, ``A population dynamics approach for the
  water distribution problem,'' \emph{International Journal of Control},
  vol.~83, pp. 1947--1964, 2010.

\bibitem{Fox2013Population-Game}
M.~J. Fox and J.~S. Shamma, ``Population games, stable games, and passivity,''
  \emph{Games}, vol.~4, pp. 561--583, Oct. 2013.

\bibitem{Franklin_jbg}
J.~Barreiro-Gomez, C.~Ocampo-Martinez, N.~Quijano, and J.~M. Maestre,
  ``Non-centralized control for flow-based distribution networks: A
  game-theoretical insight,'' \emph{Journal of The Franklin Institute}, vol.
  354, pp. 5771--5796, 2017.

\bibitem{9029756}
S.~Park, N.~C. Martins, and J.~S. Shamma, ``From population games to payoff
  dynamics models: A passivity-based approach,'' in \emph{2019 IEEE 58th
  Conference on Decision and Control (CDC)}, 2019.

\bibitem{park2018cdc}
S.~{Park}, J.~S. {Shamma}, and N.~C. {Martins}, ``Passivity and evolutionary
  game dynamics,'' in \emph{2018 IEEE Conference on Decision and Control
  (CDC)}, 2018, pp. 3553--3560.

\bibitem{9219202}
M.~Arcak and N.~C. Martins, ``Dissipativity tools for convergence to {Nash}
  equilibria in population games,'' \emph{IEEE Transactions on Control of
  Network Systems}, vol.~8, no.~1, pp. 39--50, 2021.

\bibitem{7823106}
N.~{Quijano}, C.~{Ocampo-Martinez}, J.~{Barreiro-Gomez}, G.~{Obando},
  A.~{Pantoja}, and E.~{Mojica-Nava}, ``The role of population games and
  evolutionary dynamics in distributed control systems: The advantages of
  evolutionary game theory,'' \emph{IEEE Control Systems Magazine}, vol.~37,
  no.~1, pp. 70--97, 2017.

\bibitem{9483414}
S.~Park and N.~E. Leonard, ``{KL} divergence regularized learning model for
  multi-agent decision making,'' in \emph{2021 American Control Conference
  (ACC)}, 2021, pp. 4509--4514.

\bibitem{park2023learning}
------, ``Learning with delayed payoffs in population games using
  {Kullback-Leibler} divergence regularization,'' arXiv.org, 2023.

\bibitem{Khalil:1173048}
H.~K. Khalil, \emph{{Nonlinear Systems; 3rd ed.}}\hskip 1em plus 0.5em minus
  0.4em\relax Upper Saddle River, NJ: Prentice-Hall, 2002.

\end{thebibliography}

\end{document}